\documentclass[iop]{emulateapj}
\usepackage{amsmath}
\usepackage{inputenc}
\usepackage{array}   
\usepackage{hhline} 
\usepackage[sort&compress]{natbib}

\newcommand{\myemail}{rjfarber@umich.edu}
\providecommand{\p}[2]{\ensuremath{\frac{\partial {#1}}{\partial {#2}}}}

\newcommand{\bnabla}{{\mbox{\boldmath$\nabla$}}}
\newcommand{\at}[2][]{#1|_{#2}}
\usepackage{tabularx}
\usepackage{graphicx}
\usepackage[breaklinks,colorlinks,citecolor=blue,linkcolor=magenta]{hyperref} 

\shorttitle{Cosmic ray transport and galactic winds}
\shortauthors{Farber et al.}

\begin{document}

\title{Impact of Cosmic Ray Transport on Galactic Winds}

\author{R. Farber\altaffilmark{1}, M. Ruszkowski\altaffilmark{1}, H.-Y.K. Yang\altaffilmark{2}, E.G. Zweibel\altaffilmark{3}}
\affil{Astronomy Department, University of Michigan, Ann Arbor, MI, USA; \myemail}
\affil{University of Maryland, College Park, Department of Astronomy and Joint Space Science Institute, MD, USA}
\affil{Department of Physics and Astronomy, University of Wisconsin–-Madison, Madison, WI, USA}
\altaffiltext{2}{{\it Einstein} Fellow}
\email{rjfarber@umich.edu (RF), mateuszr@umich.edu (MR), hsyang@astro.umd.edu (KY), zweibel@astro.wisc.edu (EZ)}

\begin{abstract}
The role of cosmic rays generated by supernovae and young stars has very recently begun to receive significant attention in studies of galaxy formation and evolution due to the realization that cosmic rays can efficiently accelerate galactic winds. Microscopic cosmic ray transport processes are fundamental for determining the efficiency of cosmic ray wind driving. Previous studies focused on modeling of cosmic ray transport either via constant diffusion coefficient or via streaming proportional to the Alfv{\'e}n speed. However, in predominantly cold, neutral gas, cosmic rays can propagate faster than in the ionized medium and the effective transport can be substantially larger; i.e., cosmic rays can decouple from the gas. We perform three-dimensional magnetohydrodynamical simulations of patches of galactic disks including the effects of cosmic rays. Our simulations include the decoupling of cosmic rays in the cold, neutral interstellar medium. We find that, compared to the ordinary diffusive cosmic ray transport case, accounting for the decoupling leads to significantly different wind properties such as the gas density and temperature, significantly broader spatial distribution of cosmic rays, and larger wind speed. These results have implications for X-ray, $\gamma$-ray and radio emission, and for the magnetization and pollution of the circumgalactic medium by cosmic rays.
\end{abstract}


\section{Introduction}
Galactic winds are observed ubiquitously in star-forming galaxies and significantly affect their chemical and dynamical evolution \citep{Veil05}. Galactic winds redistribute angular momentum, aiding in the formation of extended disks \citep{Broo11, Uble14}, help to produce large-scale magnetic fields in dwarf galaxies \citep{Moss17}, and pollute the intergalactic medium with metals \citep{Stei10,Boot12}.
Additionally, most galaxies are missing a large fraction of baryons compared to the cosmological average \citep{Bell03}. Models matching observed luminosity functions to simulated halo mass functions find that 20\% of the baryons are accounted for in $L_{*}$ galaxies, and that this fraction decreases rapidly for both more and less massive galaxies \citep{GuoW10}. This suggests that the efficiency of converting baryons into stars is a strong function of halo mass.
The discrepancies between halo and stellar properties, ``the missing baryons problem," constitutes an outstanding challenge in galaxy formation. Galactic winds can possibly solve the missing baryons problem by ejecting baryons out of galaxies. For galaxies more massive than $L_{*}$, active galactic nuclei likely dominate the energetics of the outflows \citep[e.g.,][]{Crot06}, while in less massive galaxies galactic winds are likely driven by stellar feedback (\citealt{Lars74}, \citealt{Chev85}, \citealt{Deke86}).

In the standard model of supernovae driven galactic winds \citep{Chev85}, thermal energy is injected into the gas, launching it ballistically and entraining denser gas as it is flung out of the galaxy. Thermally driven winds may explain the superwinds observed in starburst galaxies such as M82 (\citealt{Bust16}). However, results from high-resolution simulations demonstrated that such models may inject significant amounts of energy and launch metals out of galaxies but fail to expel a significant amount of mass into the intergalactic medium (\citealt{MacL99}; \citealt{Meli13}). Additionally, \citet{Stei10} found the kinematic features of Lyman-break galaxies best match models in which the gas velocity increases with distance to at least 100 kpc. 
This result is also difficult to reconcile with the thermal feedback model. The insufficiencies of purely thermally driven winds hint at the importance of additional stellar feedback processes, such as cosmic rays (\citealt{Boul90}; \citealt{Brei93}; \citealt{Uhli12}).

Cosmic rays can be accelerated by means of the diffusive acceleration mechanism operating in the shocks of supernova remnants (\citealt{Blan87}; \citealt{Capr15}) and in the winds from massive stars (\citealt{Byko14}). Cosmic rays exert pressure that is in rough equipartition with magnetic and dynamical pressures in the interstellar medium (\citealt{Zwei97}; \citealt{Beck01}; \citealt{Cox05}), suggesting their dynamical importance. In particular, cosmic rays can provide pressure support against self-gravitating clouds, suppressing star formation (\citealt{Jube08}; \citealt{Pfro17}). Additionally, \textit{Fermi} $\gamma$-ray observations of starburst galaxies M82 and NGC 253 imply cosmic ray energy densities roughly two magnitudes higher than in the Milky Way (\citealt{Pagl12}; \citealt{Yoas13}; \citealt{Yoas14}).

Cosmic rays escape the Galactic disk in $\sim$10 Myr \citep{Stro07}. Compared to the thermal gas, cosmic rays can be relatively free of energy losses, which together with their fast escape from the disk, suggests that cosmic rays may efficiently transport supernova energy to regions occupied by tenuous gas above the disk, which they can accelerate into a wind (\citealt{Hana13}).

Early work by \citet{Ipav75} considered the emission of magnetohydrodynamic (MHD) waves by super-Alfv{\'e}nic cosmic rays.
These waves enable cosmic rays to be coupled to the thermal gas.
A steady-state, spherically symmetric, hydrodynamic treatment by \citet{Ipav75} suggested that cosmic rays could drive outflows at rates $\gtrsim$ 1 M$_{\odot}/$yr from a typical galaxy.  \citet{Brei91} extended the work by \citet{Ipav75} by including the streaming of cosmic rays along large scale magnetic fields and found that mass outflow rates of $\sim$ 1 M$_{\odot}/$yr are possible in Milky Way-like galaxies. \citet{Ever08} further extended this work by combining cosmic ray and thermal pressure under Milky Way conditions and found that both thermal and cosmic ray pressures were essential for wind driving in the Milky Way. 

Recently, 3D numerical studies of cosmic ray winds have found that wind properties depended sensitively on the details of cosmic ray transport. This was demonstrated in both Eulerian grid hydrodynamic (\citealt{Uhli12}; \citealt{Boot13}; \citealt{Sale14}) and MHD simulations (\citealt{Hana13}; \citealt{Rusz17}) as well as unstructured moving mesh simulations (\citealt{Pakm16a}; \citealt{Simp16}; \citealt{Pakm16b}; \citealt{Pfro17}; \citealt{Jaco17}).

In predominantly cold, neutral gas, cosmic rays can propagate faster than in the ionized medium and the effective transport can be substantially larger; i.e., cosmic rays can decouple from the gas. In this work, we study the consequences of this decoupling and show that it has a significant impact on the properties of cosmic ray driven galactic winds. In Section 2, we delineate the numerical methods and treatment of physics in our simulations. In Section 3, we present our results, and in Section 4 we conclude.

\section{Methods}
We model cosmic rays with a two-fluid model (e.g., \citealt{Sale14}; \citealt{Rusz17}), in which cosmic rays take the form of an ultra-relativistic ideal fluid with an adiabatic index $\gamma_{\rm cr} = 4/3$ and the thermal gas is characterized by an adiabatic index $\gamma=5/3$. We include advection of cosmic rays, dynamical coupling between cosmic rays and the gas, and model the transport of cosmic rays relative to the gas via anisotropic diffusion of cosmic rays along magnetic field lines rather than via the streaming instability (see Section \ref{DecSubsec}). We model the effect of cosmic rays decoupling from the cold, neutral interstellar medium (ISM) via a temperature-dependent diffusion coefficient (see Section \ref{DecSubsec}). Additionally, we include star formation and feedback, self-gravity of the gas, and radiative cooling. We solve the following equations:

\begin{equation}
\p{\rho}{t} + \bnabla \cdot (\rho \textbf{u}) = -\dot{m}_{\rm form} + f_{*}\dot{m}_{\rm feed},
\end{equation}
\begin{equation}
\p{\rho \textbf{u}}{t} + \bnabla \cdot \left(\rho \textbf{u} \textbf{u} - \frac{\textbf{B}\textbf{B}}{4 \pi}\right) + \bnabla p_{\rm tot} = \rho\textbf{g} + \dot{p}_{\rm SN},
\end{equation}
\begin{align}
&\p{e}{t} + \bnabla \cdot \left[ (e + p_{\rm tot})\textbf{u} - \frac{\textbf{B}(\textbf{B} \cdot \textbf{u})}{4 \pi}\right] = \nonumber \\
&\rho\textbf{u} \cdot \textbf{g} + 
\bnabla\cdot(\boldsymbol\kappa(T)\cdot \bnabla e_{\rm cr}) - C + H_{\rm SN},
\end{align}
\begin{equation}
\p{\textbf{B}}{t} - \bnabla \times (\textbf{u} \times \textbf{B}) = 0,
\end{equation}
\begin{align}
\p{e_{\rm cr}}{t} + \bnabla \cdot (e_{\rm cr}\textbf{u}) = &-p_{\rm cr} \bnabla \cdot \textbf{u} + H_{\rm SN}  \nonumber \\
                                                         &+ 
\bnabla\cdot(\boldsymbol\kappa(T)\cdot \bnabla e_{\rm cr}), 
\end{align}
\begin{equation}
\Delta\phi = 4 \pi G \rho_{b}
\end{equation}
\noindent
where $\rho$ is the gas density, $\textbf{u}$ is the gas velocity, $\dot{m}_{\rm form}$ is a density sink term representing star formation, $f_{*} \dot{m}_{\rm feed}$ is a density source term representing stellar winds and supernovae (see Section \ref{StarStuff}), $\textbf{B}$ is the magnetic field, $p_{\rm tot}$ is the sum of the gas ($p_{\rm th}$), magnetic, and cosmic ray ($p_{\rm cr}$) pressures, $\mathbf{g} = -\bnabla\phi + \mathbf{g}_{\rm NFW}$ is the gravitational acceleration (including contributions from self-gravity of gas and stars: $-\bnabla\phi$ and dark matter: $g_{\rm NFW}$; see Section \ref{NFW}), $\dot{p}_{\rm SN}$ is the momentum injection from stellar winds and supernovae, $e = 0.5 \rho \textbf{u}^{2} + e_{g} + e_{\rm cr} + B^{2}/8\pi$ is the total energy density (where $e_{g}$ is the thermal energy density), $\boldsymbol\kappa(T)$ is the temperature-dependent diffusion coefficient (see Section \ref{DecSubsec}), $e_{\rm cr}$ is the cosmic ray energy density, $C$ is the radiative cooling term (see Section \ref{RadCoolSubsec}), $H_{\rm SN}$ is the supernova heating term (see Section \ref{StarStuff}), and $\rho_{b}$ is the total (gas and stars) baryon density.

We use the adaptive-mesh refinement MHD code FLASH4.2 \citep{Fryx00,Dube08} as extended to include cosmic rays (\citealt{Yang12}; \citealt{Yang13}; \citealt{Yang17}; \citealt{Rusz17}) to solve the above equations. We utilize the directionally unsplit staggered mesh (USM) solver \citep{LeeD09,Lee13}. The USM solver is a finite-volume, high order Godunov scheme that utilizes constrained transport to ensure divergence-free magnetic fields. 

Due to computational constraints, we employ sound speed limiting. That is, we set a ceiling to the thermal and cosmic ray energy so that the timestep does not become unfeasibly small. Specifically, we impose an upper limit on the generalized sound speed
\begin{equation}
c_{s}=\left(\frac{\gamma p_{\rm th} + \gamma_{\rm cr}p_{\rm cr}  } {\rho}\right)^{1/2}.
\end{equation}
In our fiducial runs, we limit $c_{s}$ to $c_{s,{\rm lim}}=10^{3}$ km s$^{-1}$, but we have additionally run test cases for $c_{s,{\rm lim}}=2\times 10^{3}$ km s$^{-1}$ and found no significant differences in the star formation rates (c.f., \citet{Sale14}). 

For the most expensive simulation, which employs the decoupling mechanism (Run DEC, cf. Sections \ref{DecSubsec} \& \ref{ResultsSec}), we use a raised density floor of $10^{-3}$ cm$^{-3}$, which effectively limits the Alfv\'{e}n speed. In test cases that use reduced diffusion coefficients (which are computationally easier), we found no difference in the results between runs that used the raised density floor of $10^{-3}$ cm$^{-3}$ compared to the fiducial density floor $5 \times 10^{-7}$ cm$^{-3}$.

\subsection{Gravity \label{NFW}}
We include self-gravity of baryons (gas and stars) and solve the Poisson equation using the Barnes-Hut tree solver \citep{Barn86} implemented in FLASH4.2 by Richard W\"{u}nsch (\citealt{Wuns17}). This solver allows us to use mixed boundary conditions (see Section \ref{SimSetupSubsec}). 

In addition to self-gravity, we include acceleration in the $z$-direction due to dark matter. This component of the gravitational field assumes that dark matter is distributed according to the Navarro-Frenk-White profile \citep{Nava97} and has the following form: 
\begin{equation}
g_{\rm NFW}(z) = -\frac{G M_{200} z}{|z|^{3}} \frac{ \ln(1 + x) - x/(1 + x)}{\ln(1 + c) - c/(1+c)}
\end{equation}
where $z$ is height from the midplane, $G$ is the universal gravitational constant, $M_{200}$ is the virial mass of the halo, $x = |z|c/r_{200}$, $c$ is the halo concentration parameter, and $r_{200}$ is the virial radius defined as the radius of a sphere within which the average density exceeds the critical density at redshift zero by a factor of 200; i.e., 
$M_{200}=\frac{4}{3}\pi r_{200}^{3}200\rho_{\rm crit}$. See Table 1 for the parameter values used in our simulations.

\begin{table}
  \caption{Model parameters}
  \label{tab:table1}
  \begin{center}  
  	\leavevmode
    \begin{tabularx}{0.45\textwidth}{Xl}
    \hline\hline
  Halo                  \\ \hline 
   $M_{200}^{(1)}$ &   $10^{12}$M$_{\odot}$      \\
   $c^{(2)}$ &  12                   \\ \hline
   Disk                 \\ \hline
    $\rho_{o}^{(3)}$ & 5.24 $\times 10^{-24}$ g\ cm$^{-3}$ \\
    $z_{o}^{(4)}$ & 0.325 kpc \\
   $\Sigma_{o}^{(5)}$ & 100 M$_{\odot}$\ pc$^{-2}$ \\
	$T_{o}^{(6)}$    & 10$^{4}$ K \\
	$B_{o}^{(7)} = B_{o,x}$  & 1 $\mu$G \\   \hline   
Star Formation  \\ \hline
$n_{\rm thresh}^{(8)}$ & 10 cm$^{-3}$ \\
$T_{\rm floor}^{(9)}$ & 300 K \\
$m_{*,{\rm min}}^{(10)}$ & 10$^{5}$ M$_{\odot}$ \\
$\epsilon_{\rm SF}^{(11)}$ & 0.05 \\ \hline
Stellar Feedback  \\ \hline
$f_{*}^{(12)}$ & 0.25 \\
$f_{cr}^{(13)}$ & 0.1 \\
$\epsilon_{\rm SN}^{(14)}$ & 10$^{51}$erg/(M$_{\rm sf}$c$^{2}$) \\
$M_{\rm sf}^{(15)}$ & 100 M$_{\odot}$ \\ \hline

\end{tabularx}
\end{center}

\textbf{Notes.} From top to bottom the rows contain: (1) Halo mass; (2) Concentration parameter; (3) Initial midplane density; (4) Initial scale height of the gas disk; (5) Initial gas surface density; (6) Initial temperature; (7) Initial magnetic field strength; (8) Gas density threshold for star formation; (9) Floor temperature; (10) Minimum stellar population particle mass; (11) Star formation efficiency; (12) Fraction of stellar mass returned to the ISM; (13) Fraction of supernova energy bestowed unto cosmic rays; (14) Supernova energy per rest mass energy of newly formed stars; (15) Rest mass energy of newly formed stars per supernova.

\end{table}

\subsection{Radiative Cooling \label{RadCoolSubsec}}
We implemented the \citet{Town09} exact cooling method using the \citet{Rose95} piecewise power law form of the cooling function, which extends down to a floor temperature of 300 K. The \citet{Rose95} cooling function is an approximation to the \citet{Dalg72}, and \citet{Raym76} radiative cooling functions and is given by
\begin{equation}
\Lambda(T) = \left\{\begin{aligned}  
    &0                                 &&\mbox{if                            T $<$             300} \\
    &2.2380\times 10^{-32}T^{2.0  } &&\mbox{if               300 $\leq$ T $<$            2000} \\
    &1.0012 \times 10^{-30}T^{1.5} &&\mbox{if              2000 $\leq$ T $<$            8000} \\
    &4.6240 \times 10^{-36}T^{2.867} &&\mbox{if              8000 $\leq$ T $<          10^{5}$} \\
    &1.7800 \times 10^{-18}T^{-0.65} &&\mbox{if          $10^{5}$ $\leq$ T $< 4 \times 10^{7}$} \\
    &3.2217 \times 10^{-27}T^{0.5} &&\mbox{if $4 \times 10^{7}$ $\leq$ T, }\end{aligned}\right. 
\end{equation}
where $T$ is the gas temperature in K and $\Lambda(T)$ is in the units of erg cm$^{3}$ s$^{-1}$. The above cooling function is approximately correct for gas of solar abundance, which is completely ionized at 8000 K. Unlike explicit or implicit solvers, the Townsend integration scheme is exact and does not impose restrictions on the cooling timestep. Our tests confirm that the gas temperature evolution computed using this method follows, down to machine precision, the evolution predicted analytically (see Appendix \ref{TownApp}).

\subsection{Cosmic ray decoupling from the cold interstellar medium\label{DecSubsec}}
Cosmic rays can be efficiently confined to hot plasmas by scattering off self-excited hydromagnetic waves (\citealt{Kuls69}; \citealt{Kuls05}). The relative drift speed of the cosmic rays with respect to the plasma $v_{D}$ occurs at the local Alfv{\'e}n speed $v_{A}$, unless the hydromagnetic waves are damped. \citet{Kuls71} have shown that ion-neutral damping can significantly boost the relative drift velocity of cosmic rays. Moreover, as the ionization fraction decreases, the Alfv{\'e}n speed $v_{A}\propto$ (density of ionized particles)$^{-1/2}$ increases. Both of these effects lead to the decoupling of cosmic rays from the low-temperature ISM. 

Note that turbulent damping (\citealt{Farm04}; \citealt{Laza16}; P. Holguin et al. 2018, in prep.), linear Landau damping (\citealt{Wien18}), and nonlinear Landau damping (\citealt{Kuls05}) will also increase the relative drift velocity of cosmic rays. All of these damping mechanisms (including ion-neutral damping explored in this work) dissipate cosmic ray streaming energy into heat. This will affect the equation of state, which is different for the waves (depending on the Alfv\'en Mach number), cosmic rays, and gas, an effect we neglect in the present work.

We note that even in the presence of cosmic rays in the cold ISM, the ionization fraction in most of this phase should be low because it is mostly the low-energy (tens of MeV) cosmic rays that are responsible for the ionization of hydrogen. The ionization cross section of these cosmic rays is very high \citep{Drai11} and consequently they typically do not travel far from the sites of their injection. This allows higher energy cosmic rays, that carry most of the energy of the cosmic ray fluid, to propagate in weakly ionized cold ISM where the coupling is relatively weak.

The dynamics of cosmic ray decoupling from the gas is governed by kinetic theory, yet we model cosmic rays as a fluid in order to perform simulations of tractable duration. The expectation from kinetic theory is that cosmic ray pressure and energy density tend toward constant values in space when decoupling operates (cf. \citealt{Ever11}, who also model decoupling via a large diffusion coefficient). We model the decoupling mechanism in the ``extrinsic turbulence" framework \citep{Zwei13} in which cosmic rays scatter off waves generated by turbulence driven by external sources rather than the waves generated by the streaming instability. In this model cosmic ray transport proceeds via diffusion rather than streaming, and cosmic rays exert a pressure on the gas but do not heat it \citep{Zwei17}\footnote{We have in mind here collisionless heating due to excitation and damping of Alfv{\'e}n waves; the low energy cosmic rays which ionize the gas also collisionally heat it.}.

Cosmic ray streaming and streaming heating was previously investigated in detail in \citet{Rusz17}. They found that the streaming speed (i.e., the boost factor f $>$ 1) significantly affects wind launching, while whether or not streaming heating is included does not have any noticeable impacts on the results (\citealt{Rusz17}, private communication). In this work, we focus on the influence of decoupling on the transport speed.

In order to emulate the decoupling effect in the fluid model, we adopt a simple treatment in which the parallel diffusion coefficient is amplified to large values in low-temperature regions. This has the same effect of increasing the effective transport speed in the low-temperature gas as discussed above. As such, it will smooth the gradients in cosmic ray pressure and energy density, matching the expectations from kinetic theory.

The parallel diffusion coefficient $\kappa_{||}$ can be expressed as $\kappa_{\rm Bohm}/\epsilon$, where $\epsilon$ is the ratio of the scattering frequency of cosmic rays on the waves generated by external turbulence to the gyrofrequency $c/r_{g}$ and  $\kappa_{\rm Bohm}= \frac{1}{3} r_{g} v_{p}$ is the Bohm diffusion coefficient, $r_{g}$ is the gyroradius, and $v_{p}$ is the particle speed \citep{Schl89,Enss2003}. The scattering frequency depends on the properties of the MHD turbulence on the scales comparable to $r_{g}$. Since we are working in the ``extrinsic turbulence'' model, the source of the magnetic field perturbations capable of deflecting cosmic rays is most likely compressive waves generated by the Goldreich-Sridhar cascade down to this small scale \citep{YanL02}. Frequent cosmic ray scattering on these perturbations reduces field-aligned diffusion. The scattering frequency is proportional to $\delta B^2/B^2$, where $\delta B^2$ is the power in the magnetic fluctuations corresponding to the scale equal to $r_{g}$. Thus, $\kappa_{||}\propto B^2/\delta B^2$, and we assume that when significant wave damping is present in weakly ionized regions, the amplitude of the magnetic field perturbations decreases, and the parallel diffusion coefficient is boosted. 

As an illustration that increasing the diffusion coefficient has the desired effect of flattening the cosmic ray energy density distribution, let us consider a one-dimensional and steady state form of the cosmic ray energy density equation with spatially constant velocity $v_{0}$. For the sake of simplicity, we also neglect supernova heating in this example. In this case, integration of Eq. (5) over $z$ yields
\begin{equation}
    \kappa \frac{d e_{\rm cr}}{dz} - v_{0}e_{\rm cr} =  const.
\end{equation}
Let $\kappa = \kappa_{0} e^{z/L}$. In this example, low-temperature regions correspond to large values of $z$. The solution of Eq. (10), which satisfies boundary condition $e_{\rm cr}=e_{\rm cr,0}$ at $z=0$, is
\begin{equation}
    e_{\rm cr} = e_{\rm cr,0} + \frac{\kappa_{0}}{v_{0}}\frac{de_{\rm cr}}{dz}
    \at[\bigg]{0}
    \left[ \exp \left\{\frac{v_{0} L}{\kappa_{0}}\left(1 - e^{-z/L}\right)\right\} - 1\right]. 
\label{cr_1d_soln}
\end{equation}
For $z\gg L$, $e_{\rm cr}$ approaches a constant as required to match the behavior expected from kinetic theory. Moreover, if $v_{0}L/\kappa_{0} \ll 1$, the constant value is $e_{\rm cr} \approx e_{\rm cr}(0)$ as expected. On the other hand, in the limit of $z\ll L$,
\begin{equation}
    e_{\rm cr} = e_{\rm cr,0} + \frac{\kappa_{0}}{v_{0}}  
   \frac{de_{\rm cr}}{dz}\at[\bigg]{0} 
    \left[\exp \left\{\frac{v_{0} z}{\kappa_{0}}\right\} - 1 \right]
\end{equation}
which, as expected, matches the solution of an advection-diffusion equation with a constant diffusion coefficient.

In order to capture the effect of decoupling, we implement in the code the following simple dependence of the diffusion coefficient on the gas temperature
\begin{equation}
\kappa_{\parallel}(T) = 
\begin{cases} 
1.0 \times 10^{29} \mbox{ cm$^{2}$ s$^{-1}$ if } T < 10^{4}K \\
3.0 \times 10^{27} \mbox{ cm$^{2}$ s$^{-1}$ if } T \ge 10^{4}K, 
\end{cases}
\label{DecEqn}
\end{equation}
and set $\kappa_{\perp}=3 \times 10^{26}$ cm$^{2}$ s$^{-1}$ for all temperatures. Here, $\kappa_{\parallel}$ and $\kappa_{\perp}$ denote the diffusion coefficient for cosmic ray transport parallel and perpendicular to the magnetic field, respectively. 

The adopted cold gas value of the diffusion coefficient is representative of the values inferred for the Galaxy, but the high-temperature value of the coefficient is lower. However, our volume-weighted diffusion coefficient is expected to lie in between these limits and is comparable to that adopted by \citet{Boot13}, \citet{Sale14}, and \citet{Rusz17}. This average value is somewhat smaller than the Galactic value (c.f., \citealt{Stro98}). However, the level of diffusion inferred from observations depends on the assumptions of the models used to quantify it. Specifically, diffusion coefficients derived from the GALPROP propagation model assume spatially constant diffusion coefficient and/or often assume absence of winds. Interestingly, \citet{Ptus97} (see also \citealt{Zira96}) consider an analytic model that includes both of these effects. They study cosmic ray driven winds in which they treat streaming in the diffusion approximation and include decoupling due to ion-neutral damping. They find that the level of diffusion required for consistency with the Galactic data is only $\sim$10$^{27}$ cm$^{2}$ s$^{-1}$ outside regions close to the disk midplane, and significantly higher close to the disk where decoupling operates. The vertical velocity gradient of the cosmic ray accelerated wind in their model is large and similar to the values predicted by simulations (e.g., \citealt{Sale14}). Furthermore, \citet{Joha16} (see also \citealt{Trot11}) demonstrate that propagation parameters derived from low mass isotope data differ significantly from those based on light elements, e.g., B and C, suggesting that these species probe different locations of the ISM where cosmic ray transport may occur at different rates, though their GALPROP models neglect winds. Individual supernova remnants have also been used to put constraints on the diffusion coefficient. These studies suggested that the locally measured diffusion coefficient can be around $\sim$10$^{26}$ cm$^{2}$/s to $\sim$10$^{27}$ cm$^{2}$ s$^{-1}$ when isotropic diffusion is assumed and somewhat larger $\kappa_{||}\sim 3\times 10^{27}$ cm$^{2}$ s$^{-1}$(at 1 GeV) when anisotropic diffusion is assumed (e.g., \citealt{Nava13} and references therein).

Our approximate treatment assumes that the gas is fully ionized above $10^4$ K. The ionization level changes dramatically near this temperature threshold, and we note that the results are only weakly sensitive to the exact choice of this threshold. Consequently, the exact form of the cooling function, and specifically its dependence on the gas ionization near this critical temperature, is not critical to our conclusions.

While we implement decoupling by boosting $\kappa_{\parallel}$ by a factor of thirty in regions with $T<10^{4}$ K, we experimented with larger boost factors and found that they did not significantly affect the results. Since the computational timestep is inversely proportional to the diffusion coefficient, we decided to use smaller boost factors to accelerate the computations. Tests of the anisotropic diffusion module are presented in Appendix \ref{DecoApp}.

\subsection{Star Formation and Feedback \label{StarStuff}}
We follow the star formation prescription of (\citealt{CenO92}; cf. \citealt{Task06}; \citealt{Brya14}; \citealt{Sale14}; \citealt{LiBr15}), in which stars form when all of the following conditions are simultaneously met: (i) gas density exceeds a threshold value of $n_{\rm thresh}=10$ cm$^{-3}$ (\citealt{Gned11}; \citealt{Ager13}), (ii) flow is convergent ($\bnabla \cdot \textbf{u}<0$), (iii) cooling time is smaller than the dynamical time $t_{\rm dyn} = \sqrt{3 \pi/(32G\rho_{b})}$ or the temperature is below the floor of the cooling function (see Section \ref{RadCoolSubsec}), and (iv) the cell gas mass exceeds the local Jeans mass. 

When the above conditions are met, we form a stellar population particle\footnote{\textit{Nota Bene} A stellar population particle is representative of a star cluster, not an individual star.} instantaneously. 
The mass of the particle is $m_{*} = \epsilon_{\rm SF} (dt/t_{\rm dyn}$)$\rho dx^{3}$, where $dx$ is the size of the cell in which the particle was formed and $\epsilon_{\rm SF}=0.05$ is the star formation efficiency (\citealt{Task06}; \citealt{Rusz17}). Due to computational constraints, we prevent an exceedingly large number of stellar population particles from forming by using a minimum stellar mass $m_{*,\rm {min}}$ = 10$^{5}$ M$_{\odot}$. However, even when $m_{*}<m_{*,\rm {min}}$, we still  permit stars to form with a probability of $m_{*}/m_{*,{\rm min}}$ and mass of $m_{*}=0.8\rho dx^{3}$. Whenever a stellar population particle forms, we remove its mass from the gas the moment the stellar population particle appears. 
 
We model stellar feedback by adding to the ISM: gas at the rate of $f_{*}\dot{m}$, thermal energy at the rate of $(1 - f_{\rm cr})\epsilon_{\rm SN}\dot{m}c^{2}$, and cosmic ray energy at the rate of $f_{\rm cr}\epsilon_{\rm SN}\dot{m} c^{2}$, where $\dot{m}=m_{*}\Delta t/\tau^{2}\exp(-\Delta t/\tau)$ and  $\tau =\max (t_{\rm dyn}, 10 {\rm Myr})$. In order to conserve baryons during this time-dependent feedback process, we reduce the stellar population particle mass at the rate of $f_{*}\dot{m}$. This mass exchange represents stellar mass loss due to winds and supernovae. We use $f_{*}=0.25$, $f_{\rm cr} = 0.1$, and $\epsilon_{\rm SN} = 10^{51}$erg/($M_{\rm sf}c^{2}$), where $\epsilon_{\rm SN}$ is the energy released by supernovae per rest mass energy corresponding to the mass in newly formed stars $M_{\rm sf}=100$ M$_{\odot}$ \citep{Gued11, Hana13, Rusz17}, which corresponds to a \citet{Krou01} initial mass function. Star formation and feedback parameter choices are summarized in Table 1.

\subsection{Simulation Setup \label{SimSetupSubsec}}
We simulate a slab of ISM, with box dimensions of (2 kpc)$^{2} \times$40 kpc. This domain shape and size was motivated by the results of \citet{Hill12}, who found that an extended height was crucial in establishing a realistic halo temperature distribution in their simulations. We employ periodic boundary conditions on boundaries perpendicular to the disk and ``diode" boundary conditions on boundaries parallel to the disk. Diode boundary conditions permit material to outflow from the simulation box but prevent infall (cf., \citealt{SurS16}). Note that we neglect the magnetic field amplification due to rotational shear and our simulations somewhat enhance gravitational instability since we neglect differential rotation. These caveats are important to keep in mind; however, neglecting differential rotation greatly simplifies the calculation without sacrificing its main objectives.

We use static mesh refinement throughout the duration of the simulations to maximize resolution near the disk. Although using static mesh refinement possibly underestimates shock heating of the halo gas and thus the temperature of the wind, it does not affect our main conclusions (i.e., whether winds are launched since it mainly depends on what happens close to the disk, and the relative differences of wind properties among the three transport cases presented below). We achieve a maximum resolution of 31.25 pc in the disk. Beyond $|z|>$ 2 kpc our resolution begins to degrade down to a minimum resolution beyond $|z|>$ 4 kpc of 250 pc. One of the factors limiting  the simulation timestep is the magnetic field aligned diffusion. Our maximum resolution is comparable to that achieved in \citet{Giri2016}, who also included magnetic field-aligned cosmic ray diffusion but considered a lower maximum value of the diffusion coefficient. 

We initialize a constant temperature of $10^{4}$ K and a constant magnetic field of strength 1.0 $\mu$G along the (horizontal) x-direction throughout the computational volume. The initial density distribution follows a vertical density profile given by 
\begin{equation}
\rho(z) = \begin{cases} \rho_{0} \hspace{0.1cm}\mathrm{sech}^{2}\left( \frac{z}{2 z_{0}} \right) \mbox{ $\rho(z) > \rho_{\rm crit}$} \\
                        \rho_{\rm crit}                                       \mbox{\hspace{1.75 cm}otherwise} \end{cases}
\end{equation}
where $\rho_{0}$ is the midplane density, $z_{0}$ is the scale height of the gas disk, and $\rho_{\rm crit}$ is the critical density of the universe. Normalization constant $\rho_{0}$ is obtained from the disk surface gas density $\Sigma_{0}=\int_{-20 {\rm kpc}}^{20 {\rm kpc}}\rho(z)dz$. See Table 1 for the adopted parameter values.  For a radially exponential gas distribution of $4.1\times 10^{10}$M$_{\odot}$ in baryons in a Milky Way-type galaxy with scale factor of 3.6 kpc \citep{Boot13}, the adopted gas surface density $\Sigma_{0}$  corresponds to the gas surface density averaged within a radius of $\sim$10 kpc from the Galactic Center.

The initial setup is in rough hydrostatic equilibrium; that is, the density distribution is such that there would be hydrostatic equilibrium but for the gravity due to the NFW halo. However, gravity due to the NFW halo is small compared to self-gravity near the midplane. Consequently, gas rapidly accretes onto the midplane early in the simulation. We show results for times after 40 Myr, after memory of the initial condition has been forgotten.

\section{Results and Discussion}
\label{ResultsSec}
We present results from three simulations which include self-gravity, radiative cooling, magnetic fields, star formation and feedback, and cosmic ray pressure forces, but differ in their treatment of cosmic ray transport: in run ADV cosmic rays are advected by gas motions and diffusive cosmic ray transport is neglected; in run DIF cosmic rays are advected by gas motions and additionally the anisotropic diffusion of cosmic rays along magnetic field lines with $\kappa_{\parallel}=3\times 10^{27}$ cm$^{2}$ s$^{-1}$ and $\kappa_{\perp}=3 \times 10^{26}$ cm$^{2}$ s$^{-1}$ is included; run DEC includes cosmic ray decoupling from low-temperature ISM treated via a temperature-dependent diffusion coefficient as described in Section \ref{DecSubsec}.

\begin{figure*}
  \begin{center}
    \leavevmode
        \includegraphics[width=0.7\textwidth]{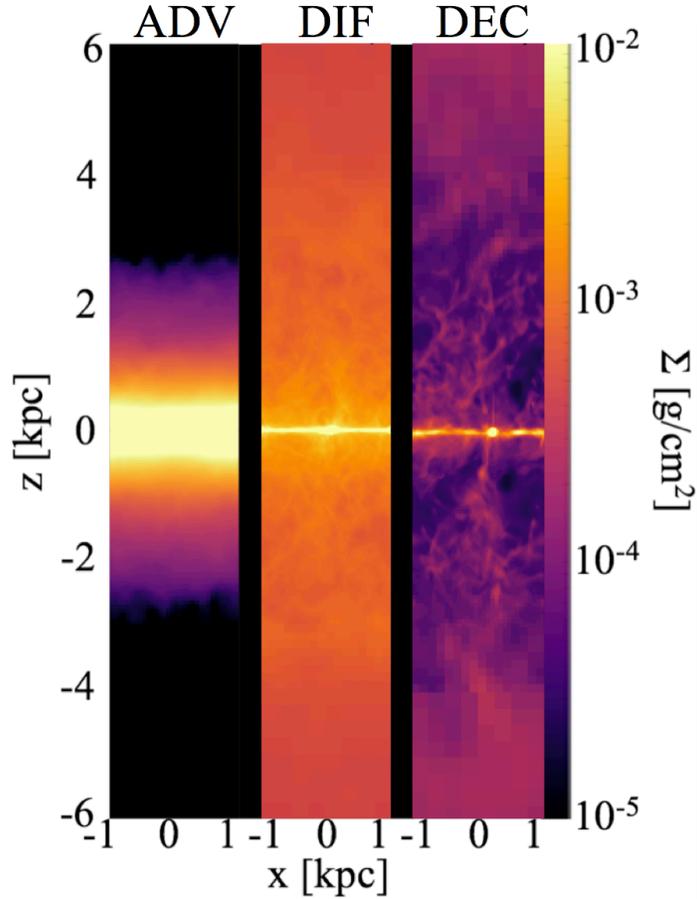}
	\caption[]{Gas mass density projections along the $y$ direction (along the midplane) for the inner $|z|<6$ kpc. The snapshots are taken at 170 Myr. Panels show three cases that correspond to different treatment of cosmic ray transport: ADV (no cosmic ray transport; left), DIF (magnetic field aligned diffusion; middle), and DEC (temperature-dependent magnetic field aligned diffusion to model cosmic ray decoupling in the cold ISM; right). Strong wind in the cases including transport, and the formation of large low-density cavities due to strong feedback in the DEC case, are evident.} 
    \label{figure1}
	\end{center}
\end{figure*}

In Figure \ref{figure1} we show the projected gas densities in the central $|z|<6$ kpc at $\sim$170 Myr for each run. In the ADV run (left panel), cosmic rays injected by supernovae are unable to efficiently drive the dense gas away from the midplane. In this case, cosmic rays remain trapped within the disk, provide additional pressure support, and thus puff up the disk. In contrast, in the DIF run (middle panel), cosmic rays are able to diffuse away from the midplane and drive the more tenuous gas located immediately above the midplane away from the disk, thus producing a wind. In the DEC run (right panel), the effects of stellar feedback are even stronger than in the DIF run -- faster cosmic ray transport out of the ISM reduces cosmic ray pressure support near the midplane, which leads to enhanced star formation rate, and thus, stronger feedback. This stronger feedback, together with the fact that cosmic rays avoid dense gas in the disk due to their decoupling from the cold ISM phase and preferentially acting on more tenuous gas, leads to the formation of hot, low-density bubbles extending up to $\pm 4$ kpc away from the midplane (see right panel in Figure \ref{figure1}). The timing of the snapshots shown in Figure \ref{figure1} corresponds to the period following the peak in star formation rate and was chosen specifically to reveal the maximum spatial extent of the hot bubbles.

Previous simulations also found that cosmic rays cannot drive winds without diffusive transport (\citealt{Jube08}, \citealt{Uhli12}, \citealt{Simp16}). However, the formation of small scale structure is resolution dependent. We performed an additional ADV simulation at 15.625 pc resolution and still found that no wind was produced. Nevertheless, it is possible that at sufficiently high resolution the thermal energy from supernovae would carve out channels in the gas density. The channels would allow the cosmic rays to escape rather than puffing up the disk. This hypothesis shall be examined by future higher resolution simulations which can better resolve high-density clumps within the disk (but this is beyond the scope of the current work).

\begin{figure*}
  \begin{center}
    \leavevmode
    \includegraphics[width=0.7\textwidth, angle=270]{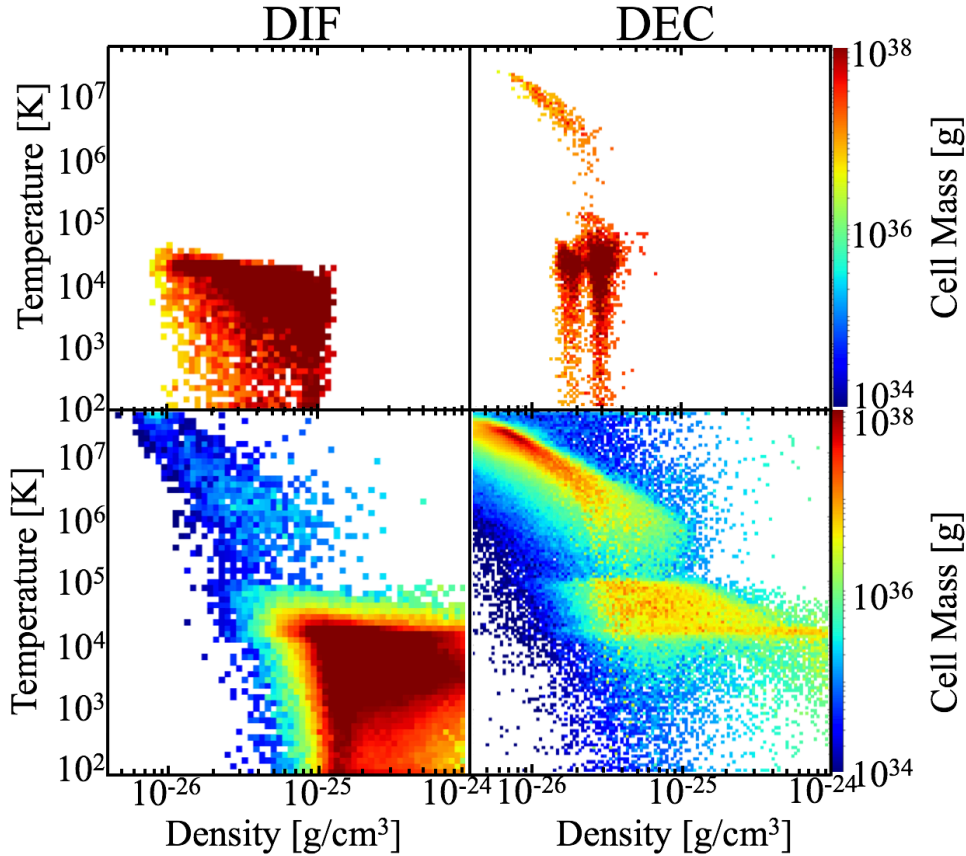}
\caption{Temperature-density phase plots at 170 Myr for $|z|<$ 4 kpc (bottom row) and $|z|>$ 4 kpc (top row). Left column corresponds to Run DIF and the right one to Run DEC.
\label{figure2}}
\end{center}
\end{figure*}
\begin{figure*}
  \begin{center}
    \leavevmode
    \includegraphics[width=0.95\textwidth]{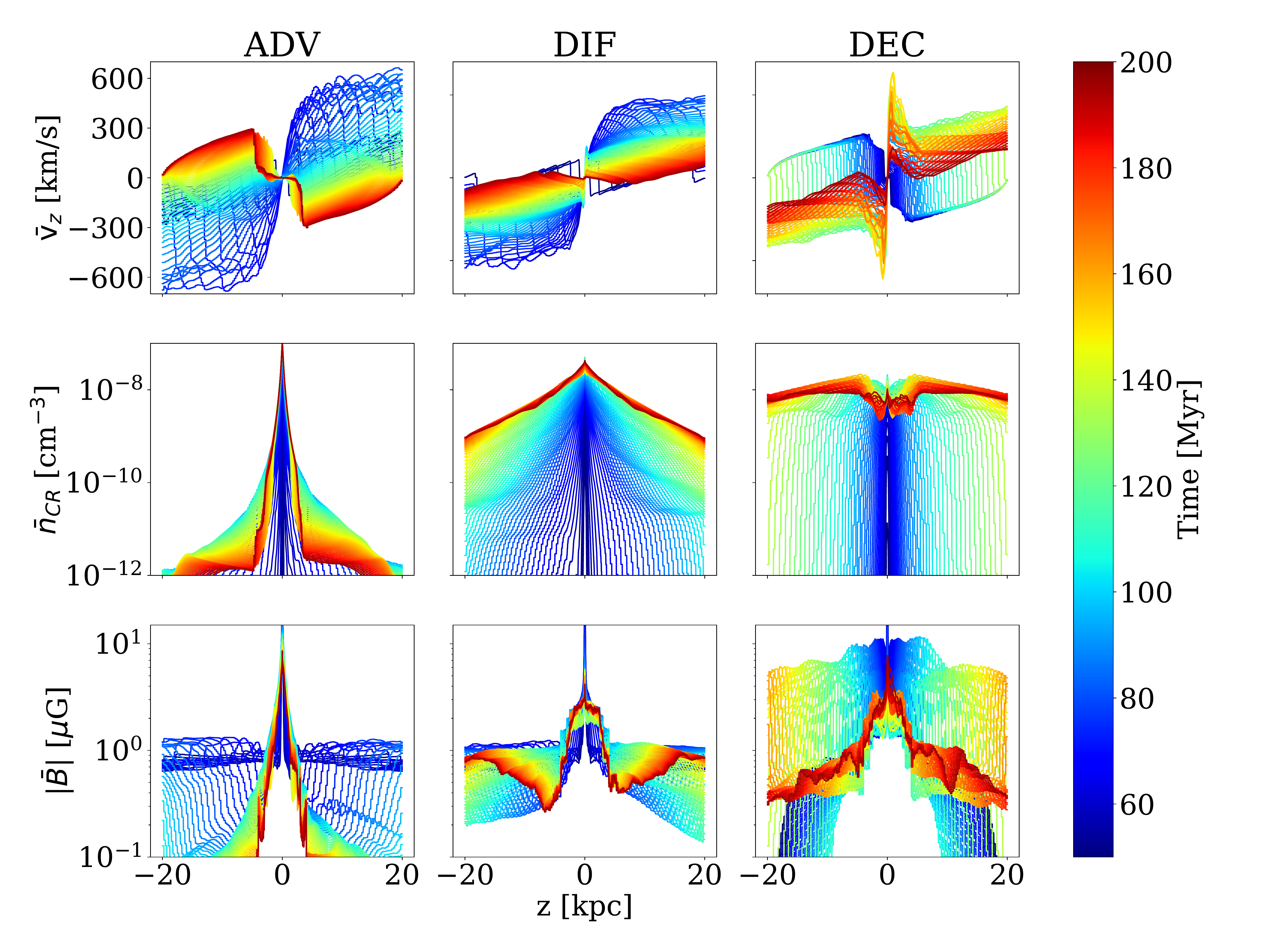}
\caption{Time series of profiles of the wind velocity (top row), the cosmic ray number density (middle row), and the magnetic field strength (bottom row) as a function of height above the midplane. All three variables are volume-weighted. 
From left to right, columns show results for the ADV, DIF, and DEC cases. The time series span the range from 50 Myr (dark blue) to 200 Myr (dark red). 
\label{figure3}}
\end{center}
\end{figure*}
\begin{figure*}
  \begin{center}
    \leavevmode
    \includegraphics[width=0.99\textwidth]{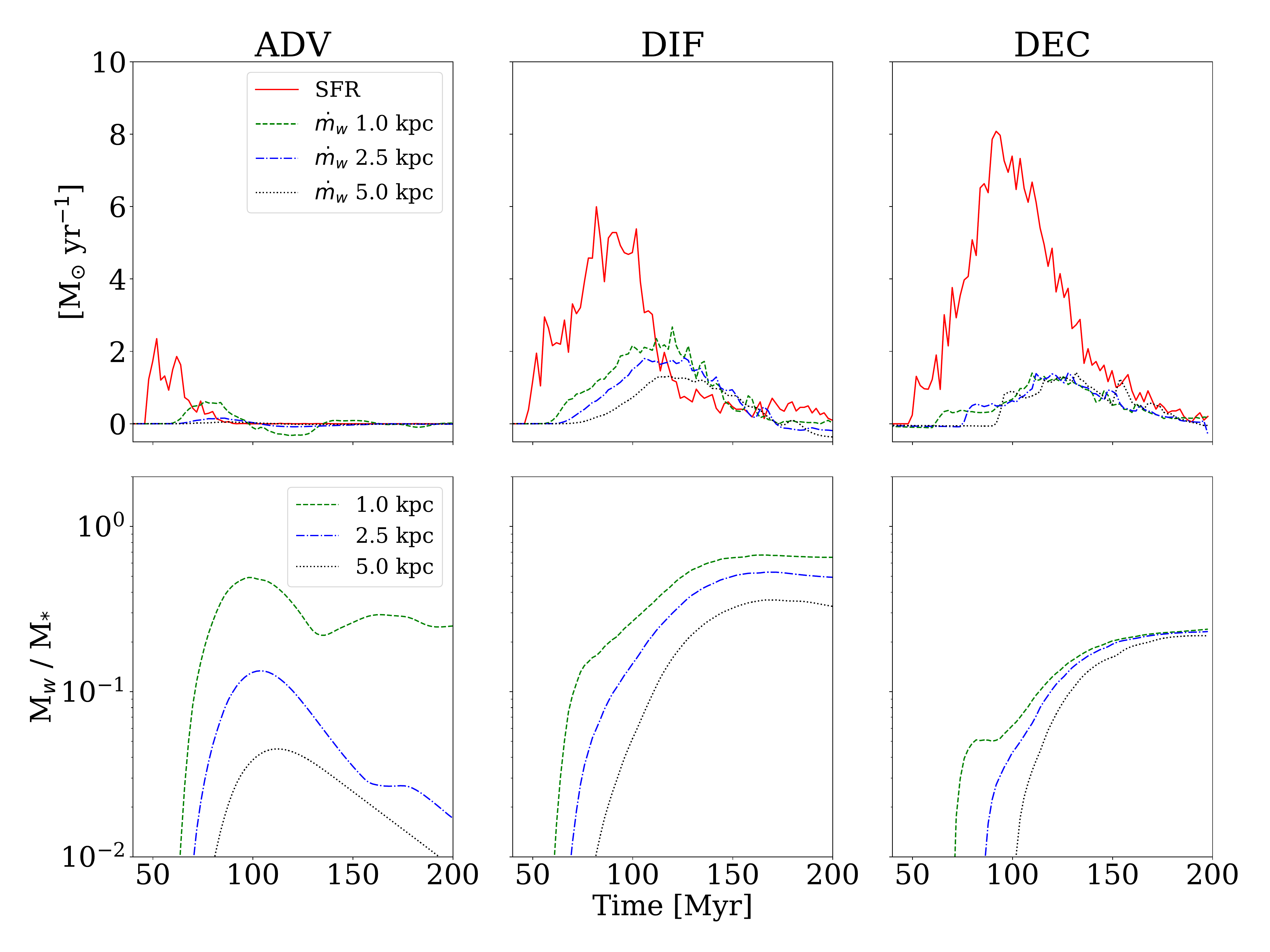}
\caption{Evolution of the mass outflow rate and star formation rate (top row) and integrated mass loading (M$_{\rm wind}$/M$_{*}$) (bottom row). Mass outflow rate is measured using surfaces parallel to the disk midplane. Curves corresponding to mass outflow rates are labeled according to the heights (measured from the disk midplane) of these surfaces (see text for details). 
\label{figure4}}
\end{center}
\end{figure*}

To better understand the impact of cosmic ray decoupling on the properties of the ISM, we next consider temperature-density phase plots. Figure \ref{figure2} shows these phase plots for $|z|<$ 4 kpc (bottom row) and $|z|>$ 4 kpc (top row). The left column shows results from the DIF run and the right from the DEC run. All panels correspond to 170 Myr. One of the most striking differences between these phase plots is the difference between the phase plots corresponding to $|z|<$ 4 kpc. These plots reveal that the low density bubbles formed in the DEC case (see right panel in Figure \ref{figure1}) are very hot ($10^{6}-10^{7}$ K). This feature is absent in the DIF case. We verified that the hot underdense bubbles are present throughout the $|z|<$ 4 kpc region rather than being confined to smaller distances from the midplane. These results also show that the gas in the DEC case is on average hotter than in the DIF case (i.e., even gas with temperatures below $10^{4}$ K is less abundant). As mentioned above, this is a consequence of enhanced stellar feedback in the DEC run. At $|z|>$ 4 kpc (top panels), the maximum gas density in the DEC case is lower compared to the DIF run. This is consistent with the gas surface density distributions shown in Figure \ref{figure1}. This could again be understood as being due to faster cosmic ray transport in the DEC case. This enhanced transport forces cosmic rays to interact with relatively less dense disk gas, which results in an outflow characterized by lower density.

Figure \ref{figure3} shows the evolution of the vertical wind velocity (top row), cosmic ray number density (middle row), and magnetic field strength (bottom row) as a function of height above the midplane. All three variables are volume-weighted. The cosmic ray number density is computed from the simulation output cosmic ray energy density by $n_{\rm cr} = [(n-4)/(n-3)]e_{\rm cr}/E_{\rm min}$ where $n=4.5$ is the slope of the cosmic ray distribution function in momentum, and $E_{\rm min}=1$ GeV is the minimum cosmic ray energy. 

From left to right, columns show results for the ADV, DIF, and DEC cases. The profiles are shown from the beginning of star formation at 50 Myr to quiescence at 200 Myr. 


We begin the discussion of Figure \ref{figure3} by considering the vertical gas velocity. In the ADV case, inflows (positive/negative ``wind" velocity at negative/positive $z$) are present for most of the simulation time. Note that regions at large heights above the disk contain very little gas in this case. This has to be contrasted with the DIF and DEC runs that are dominated by outflows. Interestingly, the wind in the DEC case is faster than in the DIF case (e.g., compare green-yellow curves near 140 Myr when the wind is about twice as fast in the DEC run) and lasts longer. Both of these effects are the result of stronger stellar feedback and the fact that it is easier to accelerate lower density gas in the DEC case. 

Let us now consider the spatial distribution of cosmic rays and magnetic fields (shown in middle and bottom row, respectively). Cosmic rays are most tightly confined to the midplane in the ADV case. This is consistent with our analysis of the gas density projections (see left panel in Figure \ref{figure1}). On the other hand, in both DIF and DEC runs the cosmic rays are much more dispersed than in the ADV case; i.e., at large $|z|$ the cosmic ray number density is much greater for most of the simulated time. Moreover, the DEC run exhibits a much wider distribution of cosmic rays than that seen in the DIF case.

The trends seen in the evolution of the cosmic ray distribution are generally reflected in the evolution of the magnetic field profiles. Specifically, the magnetic field distribution is much broader in the simulations that include cosmic ray transport (DIF and DEC) compared to the ADV case. However, the magnetic field strength is much stronger in DEC than in DIF (except at very late times). In the DEC run, cosmic ray feedback is substantially more explosive than in the DIF run, launching a strongly magnetized outflow.

While we initialize a unidirectional 1 $\mu$G magnetic field, that initial field is quickly erased. In the inflow case (initial stage of the ADV case) the field is simply accreted onto the midplane. As we use diode boundary conditions (inflow through the top and bottom boundary is not permitted), accretion leads to the reduction of the magnetic field in the regions away from the midplane. Because no wind develops in this case, the gas density above and below the midplane is also very low as can be seen in the left panel in Figure 1. Thus, we expect the results in the ADV case to be unaffected by our assumption of spatially constant magnetic field, and an  initial field decaying with the distance from the midplane should lead to very similar results. The bulk of the magnetic field amplification occurs only near the midplane as a result of turbulent motions associated with star formation and feedback. In the cases that include transport (DIF and DEC), the field is also amplified in the disk due to star formation and feedback, but following its amplification, it is expelled from the disk. During the outflow, the initial field is swept out of the simulation volume through the outer boundaries. We do not expect our assumption regarding the initial magnetic field to affect our conclusions in these cases either. Thus, the system quickly loses memory of the initial field configuration and strength. It is only in the cases that include cosmic ray transport that we expect significant magnetization of the gas at large distances from the disk at late times. This magnetization process occurs earlier in the DEC case compared to the DIF case, because the wind speed is larger in the former case. 

The above differences between the evolution of the wind velocity, gas density,  temperature, magnetic field strength, and cosmic ray number density will lead to different observational signatures. For example, we expect a stronger, spatially-extended soft X-ray emission when the decoupling mechanism operates. The presence of such emission may mitigate the problem reported by \citet{Pete15}, who found that cosmic ray driven outflows eject too little hot gas to match the soft X-ray background. Note that if streaming was additionally included, the coupled regions would be collisionlessly heated by cosmic rays, possibly producing even higher temperatures and stronger soft X-ray emission. 

Furthermore, elevated cosmic ray number densities and magnetic field strengths in the halo, combined with shorter advection times than synchrotron cooling times, suggest more extended radio emission in the DEC case than in the DIF case. However, we note that our simulations reflect the cosmic ray proton rather than cosmic ray electron distribution; cosmic ray electrons are subject to energy-dependent losses (synchrotron and inverse Compton cooling) which dominate the nonthermal radio emission. We will investigate radio spectra in future work (e.g., via a Lagrangian tracer particle approach to follow synchrotron aging of electrons co-moving with the wind). 

Finally, as the decoupling reduces the amount of time cosmic rays spend in the cold ISM phase, we expect that this mechanism would have implications for the $\gamma$-ray emission due to hadronic processes. We defer the study of these effects, and the other observational signatures, to a future publication.

In Figure \ref{figure4}, we quantify the properties of the mass flow and compare them to the star formation rates in the ADV (left column), DIF (middle column), and DEC cases (right column). Top row shows the evolution of the star formation rates (solid red lines) and the mass outflow rates computed by integrating mass fluxes through three different pairs of surfaces parallel to the disk. These planes are positioned at $\pm 1$ kpc (dashed green lines), $\pm 2.5$ kpc (dot-dashed blue lines), and $\pm 5$ kpc (dotted black lines). We find that there is essentially no outflow in the ADV case and the star formation in this case is very weak. This is consistent with the findings of a number of authors (e.g., \citealt{Sale14,Giri2016,Simp16,Rusz17}). In this case, cosmic rays are confined to the dense disk and the pressure forces they exert are too weak to expel the dense gas from the galaxy. The enhanced pressure support in the disk inhibits the collapse of cold gas clumps and thus significantly suppresses star formation.

We note that while cosmic ray advection-only simulations suppressed the star formation rate in our simulation (as likewise observed in all other works to date of which the authors are aware), it is possible that at sufficiently high resolution, the cosmic ray pressure may  

This picture is significantly altered when transport processes are included in the simulations. We find that in the DIF case, star formation is enhanced compared to the ADV case, and gas is displaced from the midplane. We observe a significant gas outflow followed by some inflow (the signature of a fountain flow). When the decoupling physics is included, the star formation rate is enhanced further and an outflow is launched, but there appears to be no inflow. Notice that not only is the star formation peak highest in this case, but the duration of the star formation episode is the longest. In DIF and DEC cases, there is a delay between the onset of star formation and the outflow. 

The star formation rate increases from ADV to DIF to DEC due to decreasing cosmic ray pressure support in the cold ISM phase caused by faster cosmic ray escape from the disk. However, including energy-dependent losses of cosmic rays could decrease the cosmic ray pressure inside dense regions in the ADV case. In such a case, cosmic rays would additionally enhance the pressure in the ambient medium, boosting the star formation rate in this case relative to the DIF or DEC cases since the Bonnor-Ebert mass goes as $P_{0}^{-1/2}$ where $P_{0}$ is the ambient pressure (\citealt{Eber55}; \citealt{Bonn56}). Thus, it is possible that a treatment including energy-dependent losses of cosmic rays would find a boosted star formation rate in the ADV case relative to the DIF or DEC cases. We will investigate the effect of energy-dependent cosmic ray transport in future work.

In the bottom row we present the evolution of the integrated mass loading factor: the ratio of the integrated mass in the wind $m_{w}$ to that in stars $m_{*}$. In the early stages of the disk evolution the gas cools and quickly settles very close to the disk midplane. Therefore, in measuring the wind mass, we delay the integration of the disk mass until 40 Myr. This allows us to exclude accretion through the set of planes positioned closest to the disk ($\pm 1$ kpc) which would appear as a negative wind mass. Thus, this approach allows us to better quantify the true amount of gas expelled to large distances from the midplane over time. 

In agreement with the findings presented in the first row, we observe that in the ADV case, the integrated mass loading factor is very low when the mass flux is measured at $\pm 5$ kpc from the disk. For smaller heights ($\pm 1$ kpc) the integrated mass loading is higher and this simply reflects the fact that the disk puffs up due to the increased pressure caused by the inability of the cosmic rays to leave the disk.

As expected, in the cases that include cosmic ray transport, the integrated mass loading factors are much larger compared to the no-transport case. Surprisingly, the level of the integrated mass loading factors is roughly comparable in both the DIF and DEC cases despite the differences in the transport physics that lead to a number of important differences in the properties of the outflows and their  observational signatures. This can be understood by the faster wind speed partially compensating for the much lower gas density in the wind for DEC compared to DIF.  The DIF run exhibits integrated mass loadings of $\sim$0.3, which is comparable to that found in other papers (e.g., \citealt{Boot13}), while the DEC run is somewhat smaller $\sim$0.2.

\section{Summary and Conclusions}
We perform simulations of cosmic ray feedback and its impact on the launching of galactic winds. In our simulations we find that cosmic ray transport is essential for driving galactic winds by cosmic rays -- in the absence of transport effects, cosmic rays alone fail to drive winds, as found in previous studies (\citealt{Jube08}, \citealt{Uhli12}, \citealt{Simp16}). 

However, a novel element of our simulations is that they incorporate the effect of the ISM temperature on cosmic ray propagation. At low temperatures, when the gas ionization fraction is low, ion-neutral friction can damp waves generated by the cosmic ray streaming instability and cosmic rays can propagate unimpeded through the ISM rather than scatter off these waves; i.e., cosmic rays are said to be decoupled from the ISM. Furthermore, low gas ionization leads to larger ion Alfv{\'e}n speed and consequently faster cosmic ray transport. Both of these effects result in faster cosmic ray transport in the cold ISM. We model this transport phenomenon by introducing enhanced diffusion in low-temperature regions. 

We note here that part of our motivation to treat decoupling via diffusion rather than streaming is to enable comparisons to previous work  that also considered diffusion. Our work generalizes previously obtained results by investigating the impact of environment-dependent diffusion. It represents the first attempt to approximate the effects of the decoupling of cosmic rays from the low-temperature plasma on wind launching and the properties of the outflow.

Our simulations focus on a patch of a galactic disk to achieve higher resolution than would be otherwise possible. These simulations address a well-posed question of how the star formation rates and wind properties are affected by the decoupling of cosmic rays from the low ionization phase of the ISM. Our specific conclusions can be summarized as follows.
\begin{enumerate}
	\item We observe the formation of low density and high temperature bubbles in the simulation that includes cosmic ray decoupling from the ISM. This raises the possibility that this case may result in enhanced soft X-ray emission from edge-on galaxies undergoing intense stellar feedback.
We suggest that the formation of these structures is due to a combination of the increase in (i) the star formation rate and (ii) the effective cosmic ray transport speed. Our simulations show that faster transport leads to the expulsion of more tenuous gas from the galaxy because cosmic rays avoid cold and dense gas clouds in the disk and preferentially act on lower density ISM.
	\item Our simulations corroborate earlier findings that cosmic ray feedback reduces star formation rates. We emphasize that this does not occur as a result of wind launching. In fact, the impact of cosmic rays on star formation is the strongest when transport processes are neglected and no wind is present. Our results are consistent with other studies in that they demonstrate a monotonic trend for the star formation to increase with the average cosmic ray transport speed. While star formation rate could be moderated by a number of model parameters, cosmic rays play a very important role in shaping the properties of the outflows and controlling star formation rates.    
\item Simulations with decoupling exhibit significantly elevated cosmic ray number densities in the halo at all times compared to the other cases. Combined with the fact the wind speed is generally faster in this case, and that the advection times are shorter while the synchrotron cooling times may be comparable to those observed in the case without decoupling, we speculate that the wider spatial distribution of cosmic rays may result in broader radio halos when the decoupling physics is taken into consideration. 
    \item Cosmic ray decoupling reduces pressure support near the galactic midplane due to faster cosmic ray escape from the cold ISM regions. This faster transport consequently leads to increased star formation rates and further injection of cosmic rays. This may have implications for hadronic losses and associated $\gamma$-ray emission.
    \item Compared to the simulations without temperature-dependent cosmic ray transport, in the simulation including decoupling, the wind speeds are larger and the wind duration is longer.   
\item The magnetic fields amplified near the midplane, that subsequently reach large distances away from the midplane, are much stronger in magnitude in the simulation including decoupling than in the case with diffusion. Additionally, since the winds are faster in the decoupling case, the dispersal of the magnetic fields occurs at earlier times in this case.  
    \item While the simulations with decoupling exhibit faster wind speeds compared to the case with diffusion, their winds are characterized by lower gas density. Consequently, wind mass loading factors -- quantified in terms of the ratio of the integrated wind mass to the cumulative mass in newly formed stars -- appear to be roughly insensitive to the physics of cosmic ray transport.
  	\item Simulations with cosmic ray transport included reveal the presence of both fountain flows and net mass loss in the wind.  
\end{enumerate}

\acknowledgments
We thank the anonymous referee for their very thoughtful report. R.F. gratefully acknowledges Hui Li, Zhijie Qu, Paco Holguin, and Nathan Goldbaum for helpful discussions. M.R. acknowledges support from NASA grant NASA ATP 12-ATP12-0017 and NSF grant AST 1715140. H.Y.K.Y. acknowledges support from NSF grant AST 1713722 and the Einstein Postdoctoral Fellowship by NASA (grant number PF4-150129). E.Z. happily acknowledges support from NSF Grant AST 1616037, the WARF Foundation, and the Vilas Trust. 
M.R. thanks the University of Maryland College Park, where part of this work was completed, for hospitality. The software used in this work was in part developed by the DOE NNSA-ASC OASCR Flash Center at the University of Chicago. Resources supporting this work were provided by the NASA High-End Computing Program through the NASA Advanced Supercomputing Division at Ames Research Center. Data analysis presented in this paper was performed with the publicly available yt visualization software (\citealt{Turk11}). We are grateful to the yt community for their support.

\appendix
\section{Exact cooling scheme test} \label{TownApp}
To test our implementation of the \citet{Town09} exact integration radiative cooling scheme, we initialized gas of constant density and temperature in a cubical box with periodic boundary conditions. This setup ensured that, despite the fact the simulation included hydrodynamics, no gas flow developed and the only quantity with time dependence was the temperature that decreased due to radiative cooling. 

To compare our implementation to an analytic solution, we simplified the problem and considered only one branch of the cooling function
\begin{equation}
\Lambda(T) = 3.2217 \times 10^{-27}T^{0.5} {\rm erg \hspace{0.1cm} cm}^{3}{\rm \hspace{0.1cm} s}^{-1}
\end{equation}
that corresponds to the uppermost temperature regime of the \citet{Rose95} cooling curve.\footnote{The full cooling function implemented in the code, and used in the simulations presented in this paper, is given by Eq. (9).} In this case, the thermal energy equation $de_{g}/dt = -n_{H}^{2} \Lambda(T)$, where $e_{g}$ is the thermal energy density, and $n_{H}$ is the hydrogen number density, has the following simple solution 
\begin{equation}
T_{f} = \left[ \sqrt{T_{i}} - \frac{3.2217 \times 10^{-27}}{3} \frac{n_{H}}{k_{B}} t \right]^{2}, 
\end{equation}
where $T_{f}$ is the temperature after elapsed time $t$, $T_{i}$ is the initial temperature, and $k_{B}$ is the Boltzmann constant.

\begin{figure}
  \begin{center}
    \leavevmode
    \includegraphics[width=0.65\textwidth]{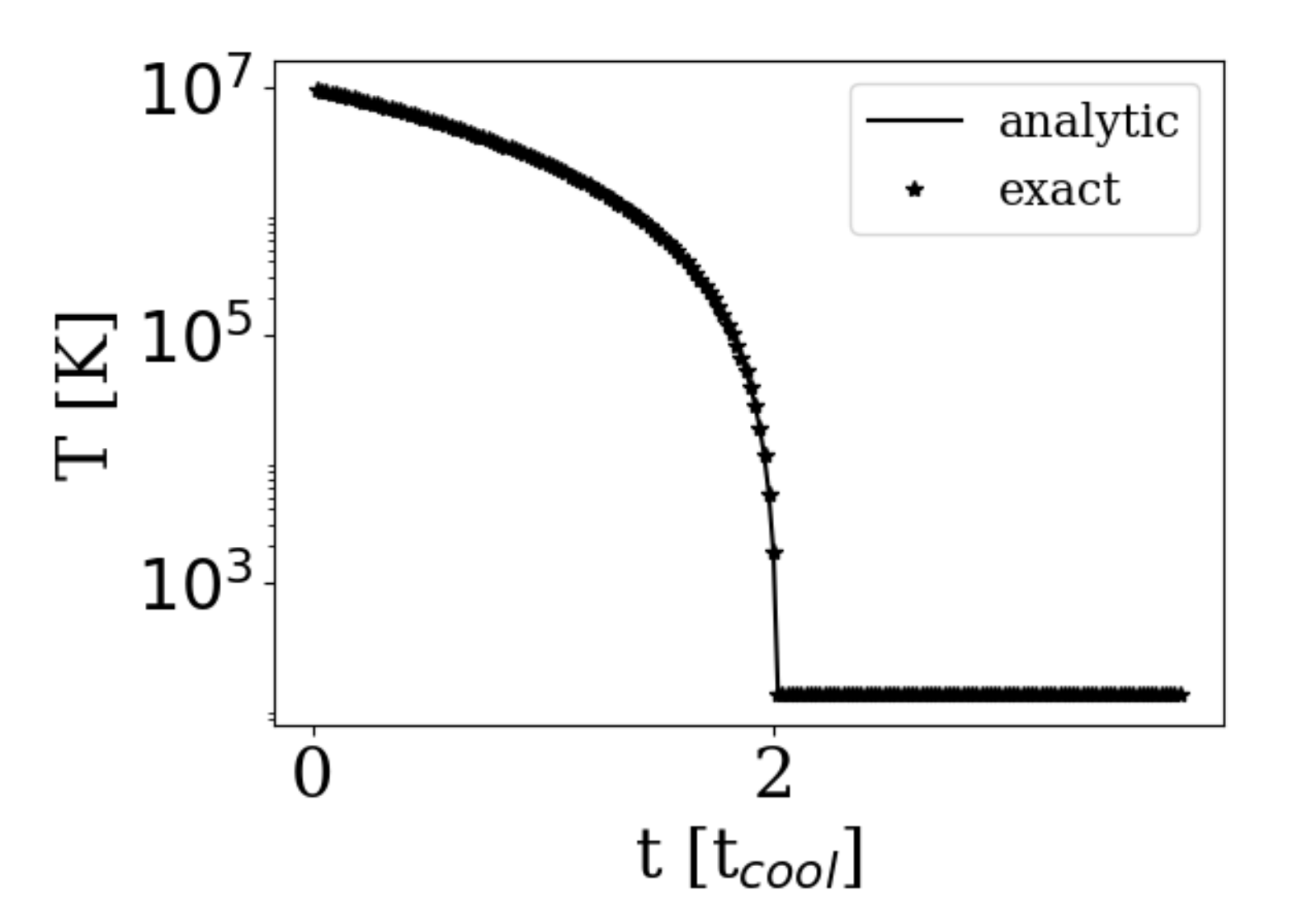}
\caption[]{Evolution of the temperature obtained using the Townsend integration method (stars) compared to the analytic result (solid line).}
\label{TownsendFigure}
\end{center}
\end{figure}

In the numerical test we set $T_{i} = 10^{7}$ K, $n_{H}=1$ cm$^{-3}$ and evolved the simulation for a few cooling times $t_{\rm cool} = e_{g}/[n_{H}^{2}\Lambda(T)]$. The agreement between the analytical result and the simulated solution obtained using the Townsend method implemented in the code was excellent (see Figure \ref{TownsendFigure}). The simulated temperature evolution agreed with the analytic result to machine precision. Importantly, the temperature did not overshoot the lowest temperature below which the cooling function was set to zero. Other methods may suffer from the overshooting problem in regions where cooling is fast and gas temperatures are low.

\section{Test of temperature-dependent cosmic ray diffusion module} \label{DecoApp}
In order to validate the cosmic ray decoupling module that we implemented, we performed the following test. We set up a simulation with a temperature-dependent diffusion coefficient $\kappa$ such that $\kappa(T)=T$ and $T(x)=1-x^{2}$. We used constant gas density and vanishing gas velocity throughout the computational domain (note that all quantities are in code units). Our spatial resolution was 64 zones in the $x$ direction and temporal resolution was $10^{-5}$ time units. The initial cosmic ray energy density was initialized according to 
\begin{equation}
e_{\rm cr}(t=0,x) = e_{\rm cr,0} + \frac{35x^4 - 30x^{2} + 3}{8},
\end{equation}
where $e_{\rm cr,0}$ is an arbitrary constant background cosmic ray energy density and the last term in Eq. (B1) is the fourth order Legendre polynomial.
These initial conditions are not isobaric. Since we were interested in testing the implementation of the diffusion module alone, we switched off hydrodynamics in this test, which allowed us to neglect pressure forces.

\begin{figure}
  \begin{center}
    \leavevmode
    \includegraphics[width=0.65\textwidth]{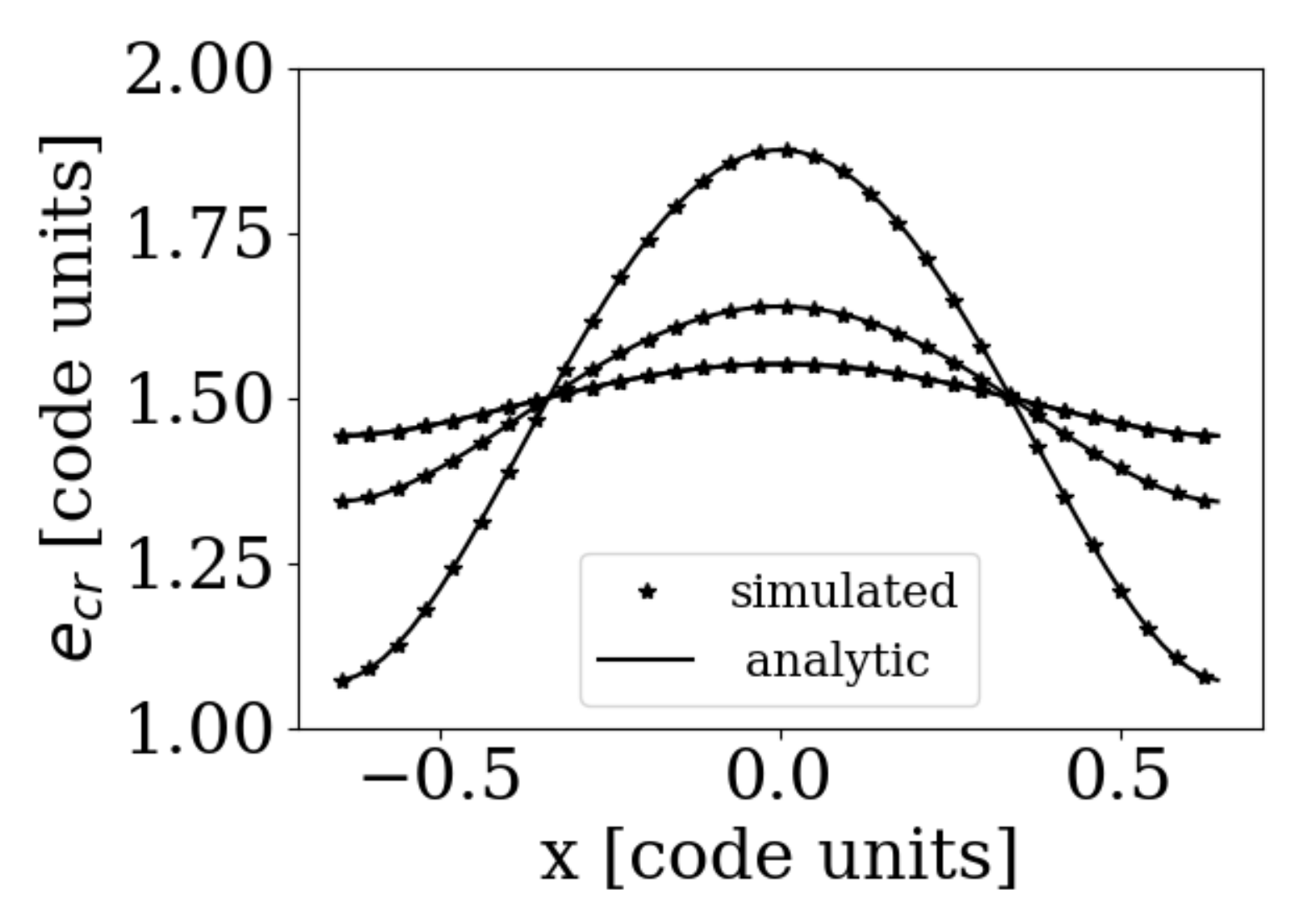}
\caption[]{The evolution of the cosmic ray energy density due to temperature-dependent diffusion. Diffusion coefficient value close to the origin is largest. Curves correspond to 0.0, 0.05, and 0.1 code time units. Flatter curves correspond to later times. The agreement between the analytic (solid line) and code solution (stars) is excellent.  \label{L4fig}}
\end{center}
\end{figure}

Using the above initial conditions, we solved the diffusion equation
\begin{equation}
\p{e_{\rm cr}}{t} = \frac{\partial}{\partial x}\left[\kappa (x) \frac{\partial e_{\rm cr}}{\partial x}\right]
\end{equation}
in a domain that was periodic in the $x$ direction. The domain consisted of 64 zones and extended from $-\sqrt{3/7}$ to $\sqrt{3/7}$. These endpoints were chosen to ensure that the initial distribution of cosmic ray energy density had a vanishing slope at the  boundaries of the periodic domain. This allowed us to eliminate jumps in energy density of cosmic rays at the boundaries. We then compared our solution to the analytic solution of Eq. (B2) given by 
\begin{eqnarray}
e_{\rm cr}(t,x) = e_{\rm cr,0} + \frac{35x^4 - 30x^{2} + 3}{8}\exp(-20t).
\end{eqnarray}
Our simulated solution was in perfect agreement with the analytical solution (see Figure \ref{L4fig}), validating the implementation of the decoupling module.\\

\clearpage

\end{document}